\documentclass[aps, prl, twocolumn, 10pt, showpacs, floatfix, superscriptaddress,longbibliography]{revtex4-1}
\usepackage{graphicx, verbatim, amsmath, amssymb,color,setspace,subfigure,braket,url}

\definecolor{darkblue}{rgb}{0.1,0.2,0.6} \definecolor{darkred}{rgb}{0.8,0.1,0.2}
\usepackage[colorlinks,citecolor=darkblue,linkcolor=darkred,urlcolor=darkblue]{hyperref}

\begin{document}
 
\title[Multiband Effects in One Dimension]{Multiband effects and the Bose-Hubbard model in one-dimensional lattices}

\author{Wei Xu}
\affiliation{Department of Physics, The Pennsylvania State University, University Park, PA 16802, USA}
\author{Maxim Olshanii}
\affiliation{Department of Physics, University of Massachusetts, Boston, MA 02125, USA}
\author{Marcos Rigol}
\affiliation{Department of Physics, The Pennsylvania State University, University Park, PA 16802, USA}

\begin{abstract}
We study phase diagrams of one-dimensional bosons with contact interactions in the presence of a lattice. We use the worm algorithm in continuous space and focus on the incommensurate superfluid Mott-insulator transition. Our results are compared to those from the one-band Bose-Hubbard model. When Wannier states are used to determine the Bose-Hubbard model parameters, the comparison unveils an apparent breakdown of the one-band description for strong interactions, even for the Mott-insulating state with an average of one particle per site ($n=1$) in deep lattices. We introduce an inverse confined scattering analysis to obtain the ratio $U/J$, with which the Bose-Hubbard model provides correct results for strong interactions, deep lattices, and $n=1$.
\end{abstract}

\maketitle

\paragraph{Introduction.}
Simplified one-band (or few-band) effective lattice models such as the Fermi-Hubbard and $t$-$J$ models for strongly interacting fermions~\cite{dagotto_review_94,imada_fujimori_98} and the Bose-Hubbard model for strongly interacting bosons~\cite{Bloch2008, Cazalilla2011} have played a central role in our understanding of the interplay between quantum fluctuations, interactions, and lattice effects in a wide range of physical systems ranging from solid-state materials to optical lattices. Unfortunately, many-body interactions make it difficult to establish the limits of applicability of such models, as well as to explore how changes in the way effective parameters are calculated extend their relevance. With those questions in mind, here we study one-dimensional (1D) systems of bosons with contact interactions in the presence of a lattice.

Effective 1D Bose systems with contact interactions are created in experiments with ultracold gases in deep two-dimensional optical lattices~\cite{moritz_stoferle_03, kinoshita_wenger_04, Cazalilla2011}. An additional (weaker) optical lattice has been used to drive superfluid Mott-insulator transitions \cite{Stoeferle2004, Haller2010, Boeris2016}. Such transitions are best understood for weak contact interactions and deep lattices, a regime that can be modeled using the one-band Bose-Hubbard model~\cite{Fisher1989, Jaksch1998}. The phase diagram of this model has been studied in great detail using a wide variety of computational techniques~\cite{batrouni_scalettar_90, elstner_monien_99, Kuehner2000, ejima_fehske_11, Carrasquilla2013}. It is well established that the phase transition driven by changing the site occupancies (incommensurate transition) belongs to the mean-field universality class, while the one with constant integer filling (commensurate transition) belongs to $(d+1)XY$ universality class~\cite{Cazalilla2011}. 

Beyond the one-band approximation, it is known that in one dimension an arbitrarily weak lattice can lead to the formation of a Mott-insulating phase in the strong interaction regime at integer fillings~\cite{Buechler2003, Haller2010}. This ``pinning'' transition is described by the (1+1) quantum sine-Gordon model~\cite{Giamarchi2003, Cazalilla2011} and has been studied very recently~\cite{Astrakharchik2016, Boeris2016}. Multiband effects have also been seen in collapse and revival experiments in three dimensions, in which they were recast into renormalized multibody interactions~\cite{Will2010,Johnson2012}, and in theoretical studies of quench dynamics \cite{Lacki_Zakrzeski_13, Lacki_Delande_13, Major_Lacki_14}. Together with other effects such as density-induced tunneling, as well as long-range interactions and tunneling~\cite{Sowinski2015}, these studies have highlighted the necessity to go beyond the standard Bose-Hubbard model to describe many experiments (for a recent review, see Ref.~\cite{Dutta2015}). 

In this Rapid Communication, we focus on the incommensurate Mott insulator transition in continuous space. To study it, we use path-integral quantum Monte Carlo simulations (QMC) with worm updates~\cite{Ceperley1995, Boninsegni2006, Boninsegni2006a}, as detailed in Ref.~\cite{Xu2015}. We compute phase diagrams obtained by changing the lattice depth while keeping the contact interaction strength constant. We consider systems with fillings of up to two bosons per site. We show that for strong contact interactions, even for deep lattices and an average of one particle per site ($n=1$), there are significant deviations from the Bose-Hubbard model predictions for the phase diagram when the parameters of that model are determined using Wannier functions. We introduce an inverse confined scattering analysis that allows one to restore the validity of the Bose-Hubbard model for strong interactions, deep lattices, and $n=1$.

\paragraph{Model Hamiltonian.}
We consider bosons with repulsive contact interactions in the presence of an external periodic potential $V_\text{ext}(x)=V_0\,\text{sin}^2(kx)$:
\begin{equation}
H=\sum_{i=1}^N\left[-\frac{\hbar^2}{2m}\frac{\partial^2}{\partial x_i^2} +V_\text{ext}(x_i)\right] +g\sum_{i<j=1}^N\delta(x_i-x_j)\,,
\label{eq:Hamilton}
\end{equation}
where $N$ is the number of particles, $V_0$ is the lattice depth, and $k=2\pi/\lambda$, with $\lambda$ being the lattice wavelength (the lattice spacing is then $a=\lambda/2$). In the absence of $V_\text{ext}(x)$, $H$ reduces to the Lieb-Liniger model, which is integrable via the Bethe ansatz~\cite{Lieb1963}. $g$ is the strength of the contact interactions, which is related to the effective 1D scattering length $a_\text{1D}$ via $g=-2\hbar^2/ma_\text{1D}$, and to the Lieb-Liniger parameter $\gamma$ via $\gamma=mg/(\hbar^2\rho)$ \cite{Cazalilla2011}. In the weak-interaction and deep-lattice limit, $H$ can be mapped onto the one-band Bose-Hubbard Hamiltonian
\begin{equation}
H_\text{BH} = -J \sum_i \left(a^\dagger_{i} a^{}_{i+1} + \textrm{H.c.} \right) 
+ \dfrac{U}{2} \sum_i n_{i} \left(n_{i} -1 \right),
\end{equation}
where $J$ is the nearest-neighbor tunneling amplitude and $U$ is the strength of the on-site repulsive interaction~\cite{Jaksch1998}. In the shallow-lattice limit, as mentioned before, $H$ reduces to the $(1+1)$ sine-Gordon Hamiltonian~\cite{Giamarchi2003, Haller2010}. This model predicts that the pinning transition occurs for $\gamma>\gamma_c\approx 3.5$ \cite{Buechler2003}. Away from those limits, exact results for phase diagrams can be obtained using QMC in continuous space~\cite{Ceperley1995, Boninsegni2006, Boninsegni2006a}. In cubic lattices, related phase diagrams have been reported in Refs.~\cite{pilati_troyer_12, nguyen_14}.

\paragraph{Indicators.} We have used two indicators to locate the transition between the superfluid and insulating phases. (i) The first is the zero-momentum Matsubara Green's function. It requires the calculation of the momentum-space Green's function $G(p,\tau)$~\cite{Capogrosso-Sansone2007}. Since the ground-state has zero total momentum, here we focus on $G(p=0,\tau)$. In the vicinity of the transition point, within the Mott-insulating phase, $G(p=0,\tau)$ decays exponentially with the imaginary time $\tau$ as $\tau\rightarrow\pm\infty$: $G(p=0,\tau)\rightarrow Z_{\pm}\text{exp}(\mp\epsilon_\pm\tau)$, where $\epsilon_{\pm}$ is the single-particle or -hole excitation energy. In the grand-canonical ensemble, $\epsilon_{\pm}$ is defined relative to the chemical potential $\mu$ with $\epsilon_{\pm}=|\mu-\mu_{\pm}|$, where $\mu_{\pm}$ gives the upper and lower boundaries of the Mott lobe. To probe the robustness of the method, we chose different chemical potentials $\mu$ and obtained essentially the same phase boundary $\mu_{\pm}$~\cite{supplement}. We also simulated different system sizes with $L/a=24,30,42, 60$ lattice sites, finding that for $L/a\gtrsim42$ finite-size effects are negligible~\cite{supplement}. All results reported from this approach are obtained from systems with $L/a=60$. (ii) The superfluid density $\rho_s = mL\braket{W^2} / (\hbar^2\beta)$~\cite{Ceperley1995}, where $\braket{W^2}$ is the winding-number estimator and $L$ is the system size. In the thermodynamic limit, $\rho_s$ is nonzero in the superfluid phase and vanishes in the Mott-insulating phase. In finite systems, $\rho_s$ satisfies a scaling relation in the critical regime~\cite{Hen2010}: $\rho_sL^{\xi/\nu}=F(|\mu-\mu_c|L^{1/\nu})$, where the critical exponent $\xi=\nu(d+z-2)$, $d=1$ is the dimension, and the correlation length and dynamical critical exponents are $\nu=1/2$ and $z=2$, respectively, for the incommensurate transition. At the critical point, $\rho_sL^{\xi/\nu}$ is independent of the system size. Thus, one can obtain the transition point from the crossing of curves for different system sizes~\cite{supplement}. 

To determine critical points, it is less computationally demanding to use $G(p=0,\tau)$ than the finite-size scaling analysis of $\rho_s$. In the former approach one only needs to do a calculation for a sufficiently large system size for one value of $\mu$ close to the phase boundary to determine $\mu_{+}$ or $\mu_{-}$, while in the latter multiple simulations with different values of $\mu$ and $L$ are needed to locate the crossing point. We then use $G(p=0,\tau)$ for constructing the phase diagrams, while the scaling of $\rho_s$ is mainly used to check results of the former approach.

\paragraph{Phase diagrams.}

\begin{figure}[!t]
 \centering
 \includegraphics[width=0.95\linewidth]{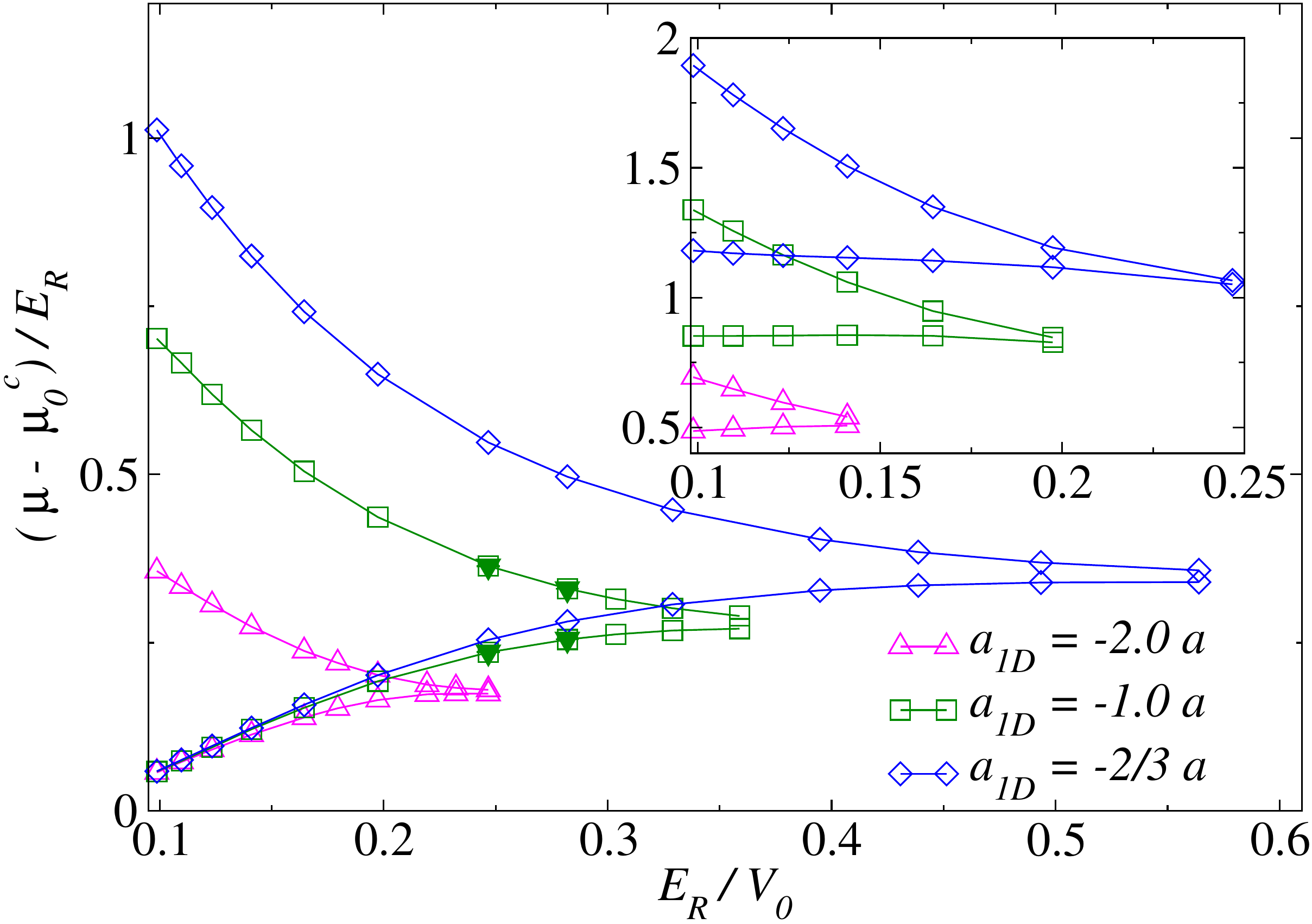}
 \caption{(Color online) Mott lobes for one particle ($n=1$, main panel) and two particles ($n=2$, inset) per site. The lobes are obtained at fixed effective 1D scattering length $a_\text{1D}/a=-2.0,\;-1.0,\;-2/3$ (for $n=1$, $\gamma=1,2,$ and 3, respectively). The points are obtained from the zero-momentum Green's function (open symbols) and the scaling of the superfluid density (solid symbols). $E_R=\hbar^2k^2/(2m)$ is the recoil energy, and $\mu_0^c$ is the chemical potential at the vacuum boundary for each value of $V_0$ (determined from the zero-momentum Green's function). Lines between symbols are to guide the eye.}
 \label{fig:pd_cont}
\end{figure}

As mentioned before, we are interested in the phase diagrams obtained by changing the lattice depth while keeping $a_\text{1D}$ (and hence $\gamma$) constant, as done in most optical lattice experiments \cite{Bloch2008, Cazalilla2011}. In Fig.~\ref{fig:pd_cont}, we report the phase diagrams obtained for three values of $a_\text{1D}$ and for Mott-insulating states with one particle ($n=1$, main panel) and two particles ($n=2$, inset) per lattice site. Note that, contrary to the usual way in which phase diagrams are reported, in Fig.~\ref{fig:pd_cont} we plot the phase boundaries in terms of $\mu-\mu_0^c$, where $\mu_0^c$ is the critical chemical potential  for the vacuum boundary as obtained in our simulations (it is independent of $a_\text{1D}$). This allows us to reduce a small chemical potential bias introduced by the finite discretization $\Delta\tau$ of imaginary time in our QMC approach. Such a bias vanishes linearly with $\Delta\tau$, and for the small but finite values of $\Delta\tau$ used in our simulations, it is negligible when chemical potential differences are reported~\cite{supplement}. In Fig.~\ref{fig:pd_cont}, open symbols depict points obtained using $G(p=0,\tau)$, while the four solid symbols for $a_\text{1D}=-a$ depict points obtained with the scaling of $\rho_s$. The latter can be seen to lead to results indistinguishable from those from $G(p=0,\tau)$. Figure \ref{fig:pd_cont} shows that, as expected, the Mott lobes grow with increasing $\gamma$. Also, the lower boundary of the $n=1$ Mott lobe becomes independent of $\gamma$ for deep lattices. All Mott lobes studied here have a finite extent because $\gamma<\gamma_c$. 

\begin{figure*}[!t]
 \centering
 \includegraphics[width=0.9\linewidth]{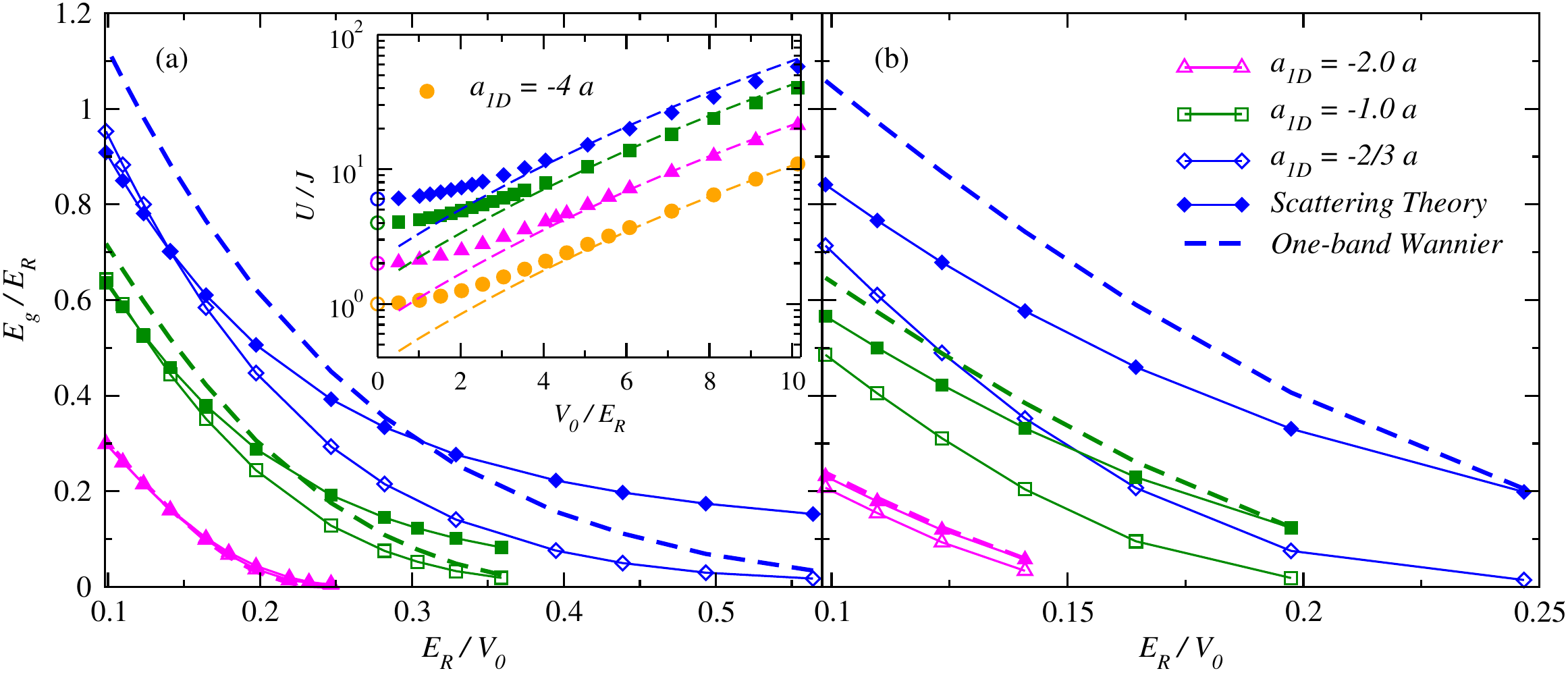}
 \caption{(Color online) Mott gaps at fillings (a) $n=1$ and (b) $n=2$ vs $E_R/V_0$ for $a_\text{1D}/a=-2.0,\;-1.0,\;-2/3$. The open symbols depict the results from QMC simulations. The dashed lines depict the Hubbard model prediction \cite{ejima_fehske_11} translated into the continuum using $U$ and $J$ as obtained from maximally localized Wannier functions \cite{Walters2013, Johnson2013}. Solid symbols depict the Hubbard model prediction \cite{ejima_fehske_11} translated into the continuum using $U/J$ from the two-particle scattering analysis explained in the text and $U$ obtained from the exact solution of two interacting bosons in a harmonic trap. Inset in (a): comparison between $U/J$ as obtained from the Wannier functions (dashed lines) and from the two-particle scattering analysis (solid symbols, consistent with those in the main panel). Open circles for $V_0/E_R=0$ show the no-lattice asymptotic limit from scattering theory.}
 \label{fig:gap_cont}
\end{figure*}

\paragraph{Gap and the one-band Bose-Hubbard model.}
Having determined the phase diagrams, we compare the predictions of the QMC simulations with those from the one-band Bose-Hubbard model. In order to minimize the number of parameters that need to be computed for the comparison, we focus on the Mott gap: $E_g=\mu_+-\mu_-$. As a first estimation, $U$ and $J$ are calculated within the one-band approximation using maximally localized Wannier states \cite{Walters2013,Johnson2013}. The Bose-Hubbard phase diagram \cite{ejima_fehske_11} is then translated into the parameters in the continuum. The results for $n=1$ and $n=2$ are presented in the main panels of Figs.~\ref{fig:gap_cont}(a) and \ref{fig:gap_cont}(b), respectively, and are compared to the QMC results derived from Fig.~\ref{fig:pd_cont}.

For $a_\text{1D}/a=-2.0$ and $n=1$ ($\gamma=1$), the gap predicted by both approaches is nearly indistinguishable for all lattice depths. The same is expected to be true for even smaller values of $a_\text{1D}$, so we focus on higher fillings and higher values of $a_\text{1D}$. For $n=2$ and $a_\text{1D}/a=-2.0$ ($\gamma=1/2$), already a small deviation can be seen between the gap obtained from QMC and the Hubbard model prediction. Striking differences, on the other hand, can be seen for $\gamma=2$ and 3, particularly for the deepest lattices and $n=2$. Counterintuitively, in Fig.~\ref{fig:gap_cont}(a), the gap predicted by the Hubbard model becomes increasingly larger than the exact one as the lattice depth is increased (the one-band prediction worsens as the lattice depth is increased). This suggests that, with increasing $V_0$, the one-band approximation leads to an increasing overestimation of $U/J$ and hence of the gap (we will come back to this point later). Hence, our results show that, for $\gamma\gtrsim 1$, the closest agreement between the exact solution and the Hubbard model prediction is obtained for the weakest lattices, for which the one-band approximation is most uncontrolled.

\paragraph{Inverse confined scattering analysis.}
When using Wannier functions, the fact that one neglects the effect of interactions is what leads to the overestimation of $U/J$ inferred from Fig.~\ref{fig:gap_cont}. In order to account for interactions in the calculation of $U/J$, so that the values of $U/J$ are the appropriate ones in the limit of deep lattices, we use an inverse scattering approach. The idea is to obtain $U/J$ by comparing the exact low-energy scattering amplitude from the Bose-Hubbard model ($a_\text{BH}$) and that of the continuum model ($a_\text{CM}$) in Eq.~\eqref{eq:Hamilton}. Our approach is different from the one followed in Refs.~\cite{Grupp2007,Buechler2010}, where scattering analyses were carried out for Feshbach resonances.
 
As in Refs.~\cite{Olshanii1998,Dunjko2001}, we define the 1D analog of the three-dimensional scattering length $a_\text{3D}$. For a translationally invariant system, the scattering wave function for two bosons with total momentum zero can be written as (in the principal domain $x_1>x_2$)
\begin{equation}
 \psi^{(2)}(x_1,x_2)=e^{i\theta(q)}\phi_q(x_1)\phi_{-q}(x_2)+\text{c.c.} \; ,
 \label{eq:fn_LL}
\end{equation}
where $\phi_q(x)$ is the plane-wave eigenfunction with positive momentum $q$ and energy $\epsilon(q)$. For any nonvanishing interaction, the scattering phase $\theta(q)$ is finite and goes asymptotically to zero as momentum $q$ approaches zero. The scattering length $a_\text{1D}$ characterizes the first nontrivial term in the Taylor expansion of the scattering phase in powers of the momentum $q$,
\begin{equation}
 a_\text{1D}=-\lim_{q\rightarrow0^+}\theta(q)/q \; .
 \label{eq:asc_def}
\end{equation}
This definition gives the effective scattering length $a_\text{1D}$ of the Lieb-Liniger model, which was introduced previously. Discretizing the Lieb-Liniger Hamiltonian, one obtains the Bose-Hubbard model and its corresponding scattering length $a_\text{BH}=-4Ja/U$~\cite{Winkler2006}.

When an external lattice is added, the propagating part of the two-body wave function has the same form as Eq.~\eqref{eq:fn_LL}, but with $\phi_q(x)$ being a Bloch-state eigenfunction and $\epsilon(q)$ being its corresponding eigenenergy. For $q$ different from half-integer multiples of the lattice wave vector $k$, $\phi_q(x)$ and $\phi_{-q}(x)$ can be shown to be linearly independent. In addition to the propagating part, there is also a localized part of the two-body wave function. It has a total energy $2\epsilon(q)$ but complex one-body energies $\zeta(\eta)$ and $\zeta(-\eta) \equiv \zeta(\eta)^{\star}$, where $\zeta(\eta)=\epsilon(q)+i\eta$ with $\eta>0$. For each $\zeta(\eta)$, one can find two linearly independent Bloch eigenstates of the one-body Hamiltonian, $H_0=-\hbar^2/(2m)(\partial^2/\partial x^2) +V_\text{ext}(x)$, labeled $\chi[k(\eta),x]$ and $\chi[-k(\eta),x]\equiv \chi[k(\eta),-x]$. They are related by a mirror reflection, where $k(\eta)$ is a Bloch momentum and $\chi[k(\eta),x]$ is a Bloch state corresponding to it. The (boson) symmetric localized part of the two-body wave function has the form (in the principal domain $x_1>x_2$)
\begin{equation}
 \chi^{(2)}(x_1,x_2,\eta)=\chi[k(\eta),x_1]\chi[k(-\eta),x_2]+\text{c.c.} \; .
 \label{eq:fn_localized}
\end{equation}
In general, there are two possible choices of the Bloch momentum $k(\eta)$ for each energy. For simplicity, we require that the imaginary part $\text{Im}[k(\eta)]>0$ for $\eta>0$~\footnote{For $\eta<0$, it can be shown that complex conjugate energies produce complex conjugate momenta, thus $k(-\eta)=k^*(\eta)$.}. We assume that the Bloch vector of the center-of-mass motion vanishes, and, as a result, the two-body scattering state is a periodic function of the center-of-mass coordinate. This leads, in turn, to the requirement that the real part of the corresponding Bloch momentum be a half-integer of $k$, i.e., $\text{Re}[k(\eta_l)]=lk/2$ for $l>0$. The full two-body scattering state can then be written as a combination of the propagating and localized components,
\begin{equation}
 \Psi^{(2)}(x_1,x_2)=\psi^{(2)}(x_1,x_2)+\sum_{l=1}^\infty A_l\,\chi^{(2)}(x_1,x_2,\eta_l) \; .
 \label{eq:twobody}
\end{equation}
$\Psi^{(2)}(x_1,x_2)$ is an eigenstate of the two-body Hamiltonian with a real eigenvalue outside of the interaction line $x_1=x_2$. Next, we introduce the relative $r=x_1-x_2$ and center-of-mass $R=(x_1+x_2)/2$ coordinates, and define $\psi_+(R,r)\equiv\phi_q(R+\frac r2)\phi_{-q}(R-\frac r2)$, of which we need $\psi_+(R)\equiv\lim_{q\rightarrow0^+}\psi_+(R,r)\big|_{r=0}$, and $\psi_+'(R) \equiv\lim_{q\rightarrow0^+} \partial_r\psi_+ (R,r)\big|_{r=0}$. The jump boundary condition~\cite{supplement} imposes the following relationship between $a_\text{CM}$ and the coefficients $A_l$ of the expansion~\eqref{eq:twobody}:
\begin{equation}
\begin{split}
\psi_+(R)a_\text{CM}=&\lim_{q\rightarrow0^+}\frac{\psi_+'(R)}{iq}a_\text{1D} \\
&-\sum_{l=1}^\infty\left[\chi^{(2)}_l(R)+a_\text{1D}\chi'^{(2)}_{l}(R)\right]\tilde{A_l} \; ,
\end{split}
\label{eq:asc_sol}
\end{equation}
where $\chi^{(2)}_l(R)=\lim_{q\rightarrow0^+}\chi^{(2)}(R+\frac r2,R-\frac r2,\eta_l)\big|_{r=0}$, $\chi'^{(2)}_{l}(R)=\lim_{q\rightarrow0^+}\partial_r\chi^{(2)}(R+\frac r2,R-\frac r2,\eta_l)\big|_{r=0}$, and
$\tilde{A}_l=\lim_{q\rightarrow0^+}iA_l/(2q)$. Since all functions of $R$ in Eq.~\eqref{eq:asc_sol} have at least a periodicity of $\pi/k$ (and are even), a way to solve that equation numerically is to use Fourier expansions. For a given number $l_\text{max}$ of terms in the sum in Eq.~\eqref{eq:twobody}, one needs to keep the same number of terms in the Fourier expansion of all the functions. As a result, one obtains an inhomogeneous $(l_\text{max}+1) \times(l_\text{max}+1)$ system of linear algebraic equations, the solution of which gives $a_\text{CM}(l_\text{max})$. We then extrapolate the results to $l_\text{max}\rightarrow\infty$ to obtain $a_\text{CM}$ (see Ref.~\cite{supplement} for more details).

In the inset in Fig.~\ref{fig:gap_cont}(a), we compare $U/J$ as obtained from the scattering and Wannier analyses for the same values of $a_\text{1D}/a$ as in the main panels, as well as for $a_\text{1D}/a=-4$. For weak interactions ($a_\text{1D}/a\lesssim-2$) and deep enough lattices ($V_0/E_R\gtrsim5$), one can see that the results from both approaches are nearly indistinguishable. On the other hand, for $a_\text{1D}/a>-2$ one can see that the Wannier analysis increasingly overestimates $U/J$ as the lattice depth and $a_\text{1D}$ are increased. This can be intuitively understood because interactions make the wave function of two particles in a site increasingly rigid to deformation as the lattice depth is increased. In that strongly interacting regime, the Wannier function calculations are not reliable even for deep lattices. As expected, independent of the value of $a_\text{1D}$, $U/J$ from the inverse scattering analysis strongly deviates from the Wannier predictions for weak lattices. In the limit of vanishing lattice depth, the former predicts $U/J=-4a/a_\text{1D}$. 

In the main panels of Fig.~\ref{fig:gap_cont}, we show the Bose-Hubbard model predictions for the phase diagram translated into the continuum using the scattering analysis results for $U/J$ and $U$ obtained using the ground-state wave function of two interacting bosons in a harmonic trap \cite{busch_98,supplement}. For $n=1$ [Fig.~\ref{fig:gap_cont}(a)] and $\gamma=1$, there is almost no visible difference from the Wannier results. On the other hand, for $\gamma=2$ and deep enough lattices, the Hubbard model with the improved values of $U/J$ and $U$ correctly predicts the value of the gap. For the deepest lattices for $\gamma=3$, we find a small deviation between the gaps predicted by QMC and by the Bose-Hubbard model. Its most likely origin is the failure of the harmonic potential to correctly predict $U$. Further studies are needed to find more accurate ways to determine $U$.  For $n=2$ and $a_\text{1D}/a\gtrsim-2$, the Bose-Hubbard results are clearly inadequate independent of how $U/J$ and $U$ are calculated.

In conclusion, our {\it ab initio} calculation of the phase diagram of 1D bosons in an optical lattice shows that, for $n\lesssim1$, the one-band Bose-Hubbard model remains useful as an effective theory far into the multiband regime, provided that its parameters are properly renormalized to account for the contributions of the excited bands. Here, the renormalized parameters are obtained from an inverse confined scattering analysis. For $n\gtrsim2$ and $\gamma>1/2$, our results highlight the need for an effective theory more refined than the traditional one-band Hubbard model. Having found experimentally relevant regimes \cite{Boeris2016,xia_zundel_15_98} in which the traditional Bose-Hubbard model fails, our study is a first step in the needed exploration of beyond-Bose-Hubbard-model physics in 1D lattices in the presence of strong interactions and/or high fillings.

\begin{acknowledgments}
 This work was supported by the U.S. Office of Naval Research, Award No. N00014-14-1-0540. The computations were performed at the Institute for CyberScience at Penn State. We thank N. Prokof'ev for providing the worm algorithm code used in the calculations, S. Ejima for sharing with us the critical values of $U/J$ reported in Ref.~\cite{ejima_fehske_11}, and J. Carrasquilla, V. Dunjko, A. Del Maestro,  Z. Yao and C. Zhang for discussions.
\end{acknowledgments}
 
\bibliography{MottInsulator1D}

\section{Supplemental Material}

\paragraph{Gap from Matsubara Green's Function.}

\begin{figure}[!t]
 \centering
 \includegraphics[width=0.84\linewidth]{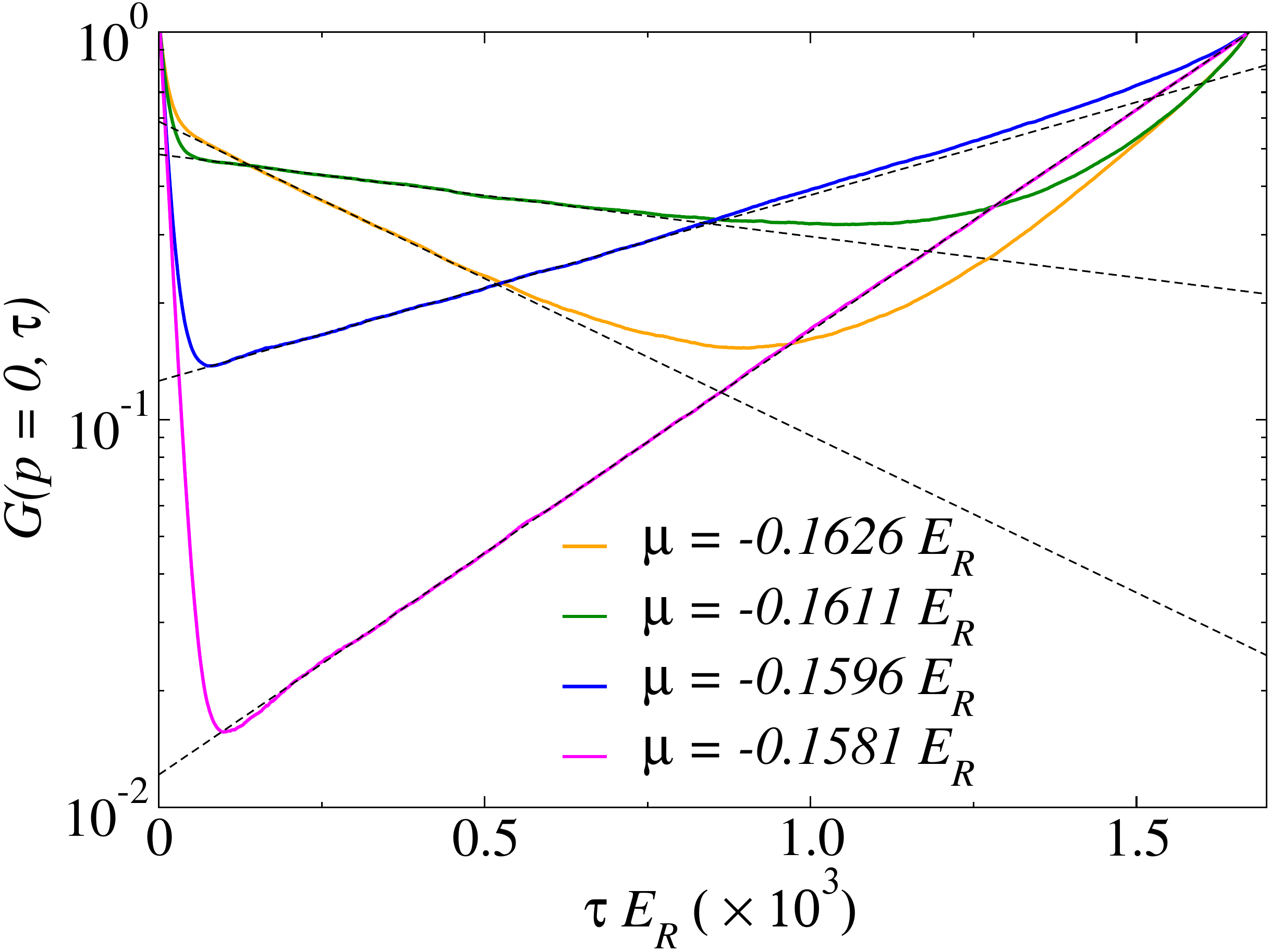}
 \vspace{-0.1cm}
 \caption{(Color online) $G(p=0,\tau)$ vs $\tau$ for different values of the chemical potential near the lower phase boundary of the Mott lobe with filling $n=1$. The lattice depth is $V_0/E_R=3.55$, the scattering length is $a_\text{1D}/a=-1$, and in all cases the system has $L/a=42$.}
 \label{fig:GG_mu}
\end{figure}

As mentioned in the main text, in order to obtain the critical chemical potential one can use the zero-momentum Green's function. However, there are two issues that need to be addressed. The first one is, given a particular system size, how sensitive the result for $\mu_+$ or $\mu_-$ (depending on the boundary of interest) is from the specific value of $\mu$ selected close to the boundary but within the Mott lobe. To check for this, we did simulations with different input chemical potentials but the same value of the lattice depth $V_0/E_R=3.55$ and scattering length $a_\text{1D}/a=-1$. The results are reported in Fig.~\ref{fig:GG_mu} for a system with $L/a=42$. For all chemical potentials selected close to the Mott-insulator--superfluid boundary, one can identify the exponential behavior of the Matsubara Green's functions. From the fit of the exponent $\epsilon_-=\mu-\mu_-$, we obtain essentially the same value $\mu_-/E_R=-0.1607\pm0.0001$ for all values of $\mu$ used.

\begin{figure}[!b]
 \centering
 \includegraphics[width=0.85\linewidth]{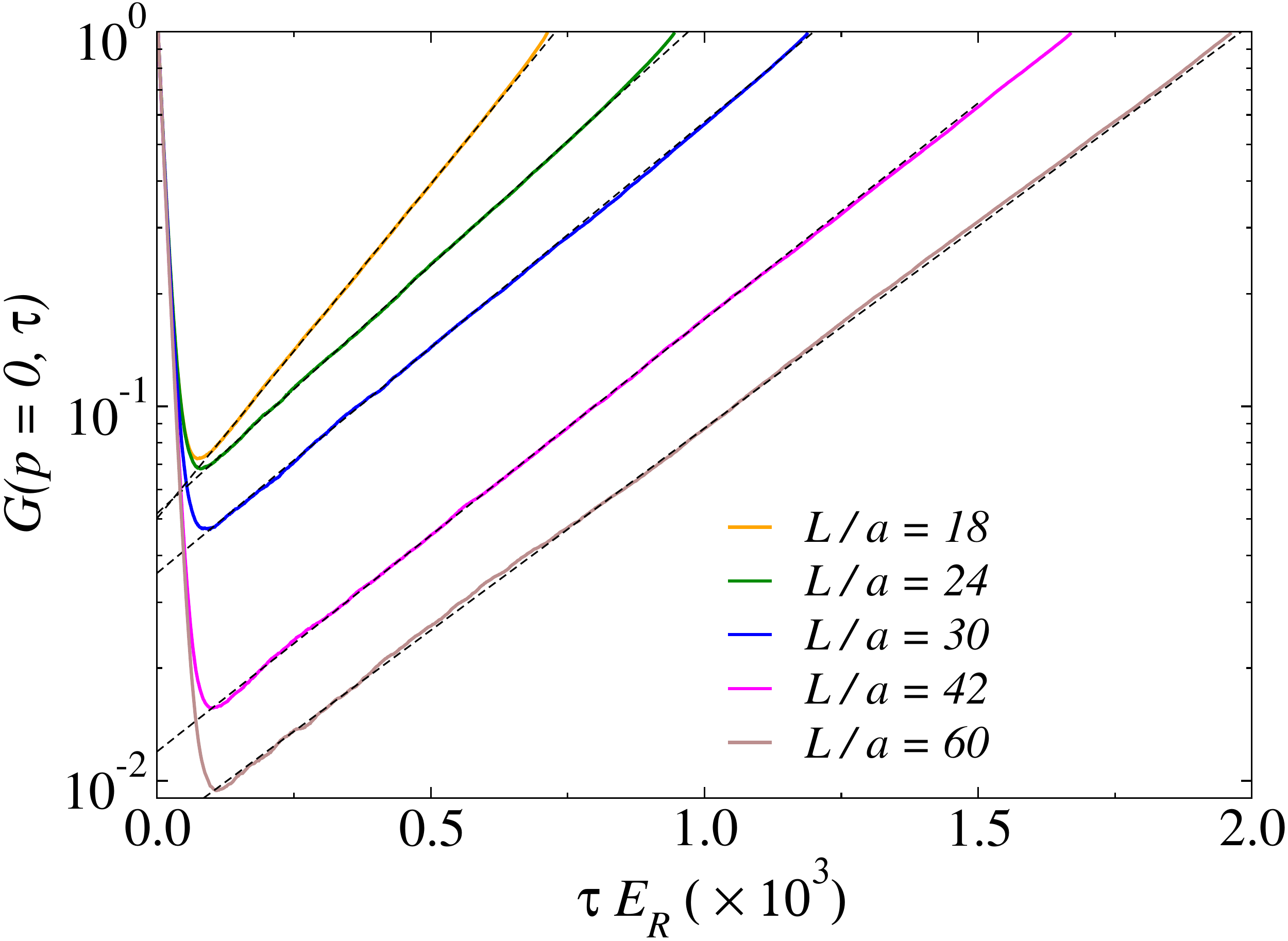}
 \vspace{-0.1cm}
 \caption{(Color online) $G(p=0,\tau)$ vs $\tau$ in systems with different number of particles and sites, but such that $n=1$. The lattice depth is $V_0/E_R=3.55$, the scattering length is $a_\text{1D}/a=-1$, and the chemical potential is $\mu=-0.1581E_R$.}
 \label{fig:GG_L}
\end{figure}

The second issue that needs to be addressed in the effect of the system having a finite extent. Figure~\ref{fig:GG_L} depicts results for the same lattice depths and scattering length as Fig.~\ref{fig:GG_L}, $\mu=-0.1581E_R$, and for different number of bosons. The slope of the curves is essentially the same for the largest system sizes, which makes apparent that finite-size effects are negligible for $L/a\gtrsim42$. All results in the main text were obtained for $L/a=60$.

\begin{figure}[ht]
 \centering
 \includegraphics[width=0.85\linewidth]{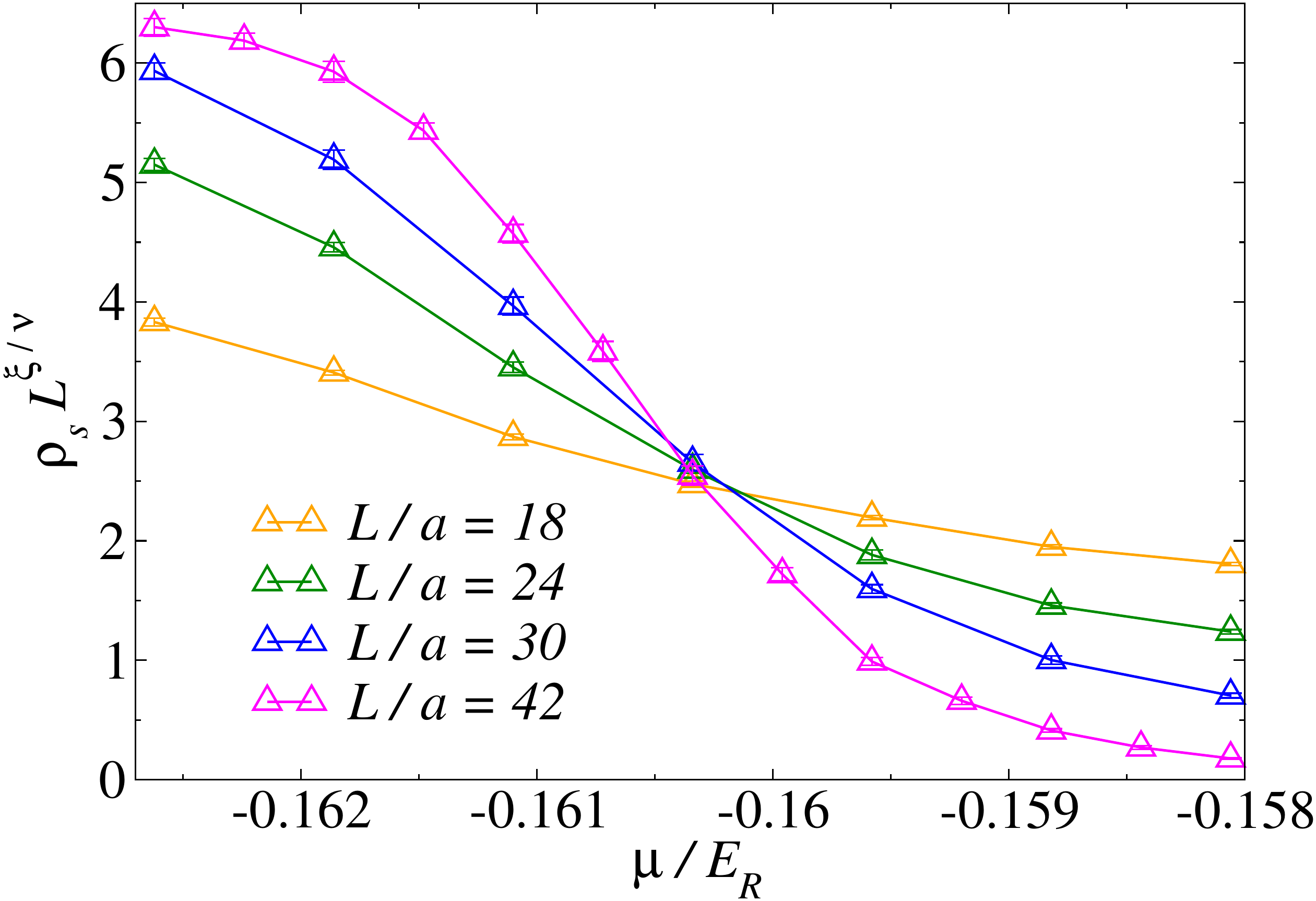}
 \vspace{-0.2cm}
 \caption{(Color online) Finite-size scaling analysis of the superfluid density for the incommensurate transition at the lower boundary of the first lobe ($n=1$). The lattice depth is $V_0/E_R=3.55$ and the scattering length is $a_\text{1D}/a=-1$.}
 \label{fig:fsc_ns}
\end{figure}

\paragraph{Finite-Size Scaling of the Superfluid Fraction.}
As also mentioned in the main text, one can obtain the critical chemical potential for a given lattice depth and strength of the contact interaction by doing a finite-size scaling analysis of the superfluid density $\rho_s$~\cite{Hen2010}. Figure~\ref{fig:fsc_ns} shows an example of such an analysis for the $n=1$ lower boundary, for a lattice depth $V_0/E_R=3.55$ and a scattering length $a_\text{1D}/a=-1$ (as in Figs.~\ref{fig:GG_mu} and \ref{fig:GG_L}). For sufficiently large systems sizes, the transition point and its error can be obtained by taking the mean value and the standard deviation, respectively, of the series of crossing points between $\rho_s$ for system size $L_i$ and $L_{i+1}$, where $i=1,2,...$ labels the system sizes simulated. For Fig.~\ref{fig:fsc_ns}, this analysis gives $\mu_-/E_R=-0.1602\pm0.0002$. This result is consistent with the value obtained using the Matsubara Green's function.\\

\paragraph{Scaling of the Phase Boundaries from the Finite Discretization $\Delta\tau$.}

In our QMC approach, imaginary time needs to be discretized. For translational invariant systems, this discretization introduces an error in the calculated physical observables that vanishes as $\Delta\tau^2$~\cite{Ceperley1995}. When calculating the critical chemical potential for the superfluid to Mott insulator transition in the presence of an optical lattice, the error introduced by the imaginary time discretization vanishes linearly with $\Delta\tau$. 

\begin{figure}[!t]
 \centering
 \includegraphics[width=1.0\linewidth]{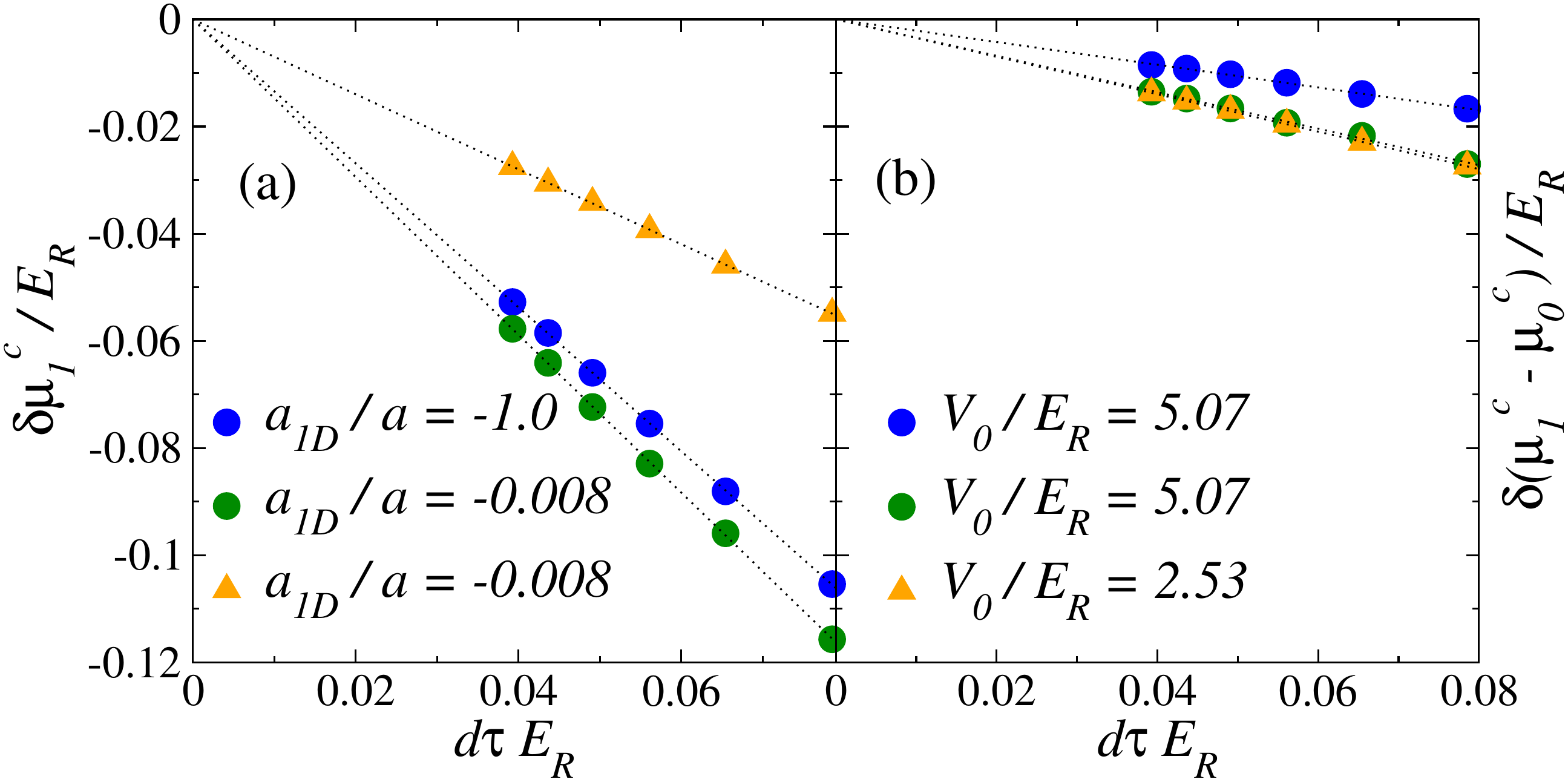}
 \vspace{-0.6cm}
 \caption{(Color online) Scaling of the critical chemical potential at the lower phase boundary of the $n=1$ Mott lobe as a function of the imaginary time discretization $\Delta\tau$, for the Tonks-Girardeau limit $a_\text{1D}/a=-0.008$ and for $a_\text{1D}/a=-1$ ($\gamma=2$). Panel (a) shows the scaling of $\delta\mu^c_1$ and panel (b) shows the scaling of $\delta(\mu^c_1-\mu^c_0)$. See the text for the definitions of $\delta\mu^c_1$ and $\delta(\mu^c_1-\mu^c_0)$.}
 \label{fig:dt_mu}
\end{figure}

In Fig.~\ref{fig:dt_mu}(a), we plot how the error for determining the lowest boundary of the $n=1$ Mott insulator in the Tonks-Girardeau limit ($|a_\text{1D}|/a\ll1$) scales with $\Delta\tau$ for $L/a=60$ (the system size used throughout the text). We define $\delta\mu^c_1\equiv \mu^c_1(\Delta\tau)-\mu^c_1(\Delta\tau=0)$, where $\mu^c_1(\Delta\tau=0)$ is obtained from a linear fit to the data. The linear scaling is apparent in the plots. We have compared $\mu^c_1(\Delta\tau=0)$ to the prediction of a band-structure calculation for free fermions $\mu^c_f$. Since the two should agree as $\gamma\rightarrow\infty$, this comparison provides a way check the correctness of $\mu^c_1(\Delta\tau=0)$. We have found that $\mu^c_f-\mu^c_1(\Delta\tau=0)<0.004E_R$, in all cases analysed. Deeper lattices lead to larger errors, as one needs prohibitively smaller values of $\Delta\tau$ to reach the appropriate scaling regime.  Figure~\ref{fig:dt_mu}(a) also shows results for $\gamma=2$ ($a_\text{1D}=-1.0a$) and $V_0=5.07E_R$. The scaling is still linear but, when compared with the Tonks-Girardeau limit, the magnitude of the prefactor is smaller, i.e., the error decreases as the contact interaction strength decreases.

Since doing such an scaling analysis for all points in the phase diagrams reported in the main text is computationally prohibitive for us, instead of reporting $\mu^c_1(\tau)$ (or any other boundary), we have reported differences with respect to the chemical potential at the vacuum boundary obtained also from QMC simulations: $\mu^c_0(\tau)$. In Fig.~\ref{fig:dt_mu}(b) we show how $\delta(\mu^c_1-\mu^c_0)\equiv [\mu^c_1(\Delta\tau)-\mu^c_0(\Delta\tau)]-[\mu^c_1(\Delta\tau=0)-\mu^c_0(\Delta\tau=0)]$ scales with $\Delta\tau$. One can see that while the scaling is still linear, and nearly independent of the lattice depth, the error due to a finite value of $\Delta\tau$ in $\mu^c_1(\Delta\tau)-\mu^c_0(\Delta\tau)$ is significantly smaller (and negligible in the scale of the plots in the main text) than the one in $\mu^c_1(\Delta\tau)$, specially for deep lattices. The error is further reduced with reducing the contact interaction strength. In Fig.~\ref{fig:dt_mu}(b), the error for $a_\text{1D}=-1.0a$ is almost one-half of the error in the Tonks-Girardeau limit.

\paragraph{General Properties of the Two-body Scattering Wavefunction.}

$\Psi^{(2)}(x_1,x_2)$ is symmetric with respect to particle permutation (because of bosonic statistics),
\begin{equation}
 \Psi^{(2)}(x_1,x_2)=\Psi^{(2)}(x_2,x_1) \; .
\end{equation}
Thus one has that:
\begin{equation}
\begin{split}
&\psi^{(2)}(x_1,x_2)= \\
  &\begin{cases}
      e^{i\theta(q)}\phi_q(x_1)\phi_{-q}(x_2)+\text{c.c.}\;, & x_1-x_2\rightarrow+\infty \\
      e^{i\theta(q)}\phi_q(x_2)\phi_{-q}(x_1)+\text{c.c.}\;, & x_1-x_2\rightarrow-\infty
   \end{cases}
\end{split}
\label{eq:fn_LL_sub}
\end{equation}
In Eq.~(1) in the main text, $\hat{H}$ is invariant under a global translation by integer multiples of $a$ and under inversion. This means that in the zero-momentum sector:
\begin{equation}
\begin{split}
&\Psi^{(2)}(x_1+a,x_2+a)=\Psi^{(2)}(x_1,x_2) \; , \\
&\Psi^{(2)}(x_1,x_2)=\Psi^{(2)}(-x_1,-x_2) \; .
\end{split}
\end{equation}
Equation~\eqref{eq:fn_LL_sub} satisfies these two conditions. Also, note that from Eq.~\eqref{eq:fn_LL_sub} it follows that $\Psi^{(2)}(x_1,x_2)$ is real.

The contact interaction between particles imposes a jump condition on the wavefunction. In the bosonic case, it leads to the following boundary condition in the principal domain $x_1>x_2$
\begin{equation}
 \left.\partial_r\Psi^{(2)}(R+\frac{r}2,R-\frac{r}2)\right|_{r\rightarrow0^+}=-\frac{1}{a_\text{1D}}\Psi^{(2)}(R,R) \; .
 \label{eq:bc}
\end{equation}
where, as in the main text, we have used the relative $r=x_1-x_2$ and center-of-mass $R=(x_1+x_2)/2$ coordinates.

\paragraph{Calculating Scattering Length.}

\begin{figure}[!b]
 \centering
 \includegraphics[width=0.84\linewidth]{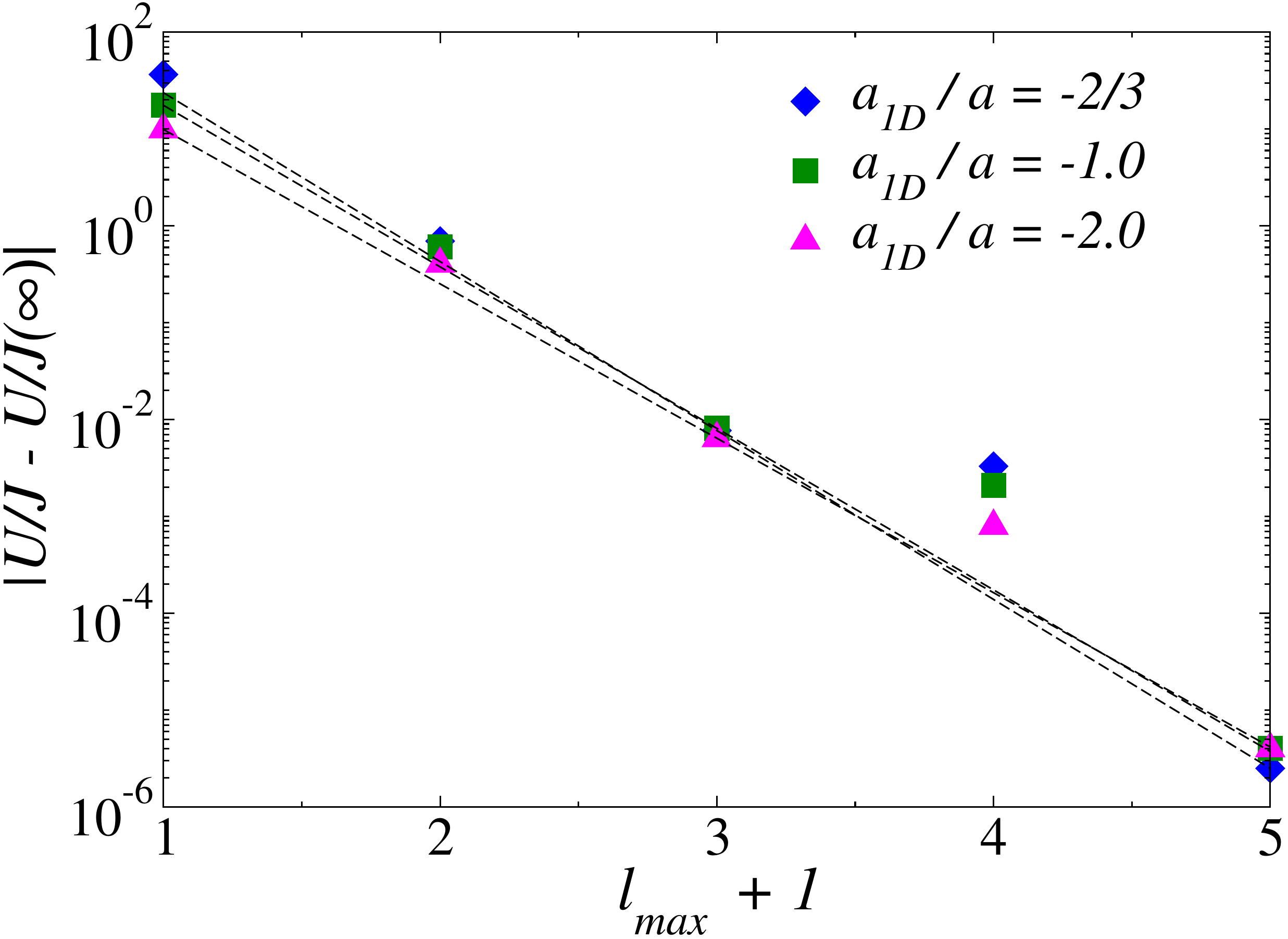}
 \vspace{-0.2cm}
 \caption{(Color online) Examples of exponential fits to the solution of $U/J$ as a function of $l_\text{max}$ for different lattice depths and for a scattering length $a_\text{1D}/a=-1.0$. The three values of $U/J(\infty)$ in the figure were obtained from exponential fits to the points with $l_\text{max}+1=1$, 3, and 5.}
 \label{fig:scatt_exp}
\end{figure}

As mentioned in the main text, Fourier expanding all functions in Eq.~(7) leads to an inhomogeneous $(l_\text{max}+1) \times(l_\text{max}+1)$ system of linear algebraic equations:
\begin{equation}
 \hat{M}\vec{V}=\vec{B} \; ,
\end{equation}
where
\begin{equation}
\begin{split} &
    \begin{split}
    M_{m,n}=&\psi_+(m)\delta_{n,0}+ \\
    &[\chi^{(2)}_n(m)+a_\text{1D}\chi'^{(2)}_{n}(m)](1-\delta_{n,0}) \; ,
    \end{split}  \\
&V_n=a_\text{CM}(l_\text{max})\delta_{n,0}+\tilde{A}_n(1-\delta_{n,0}) \; , \\
&B_m=\lim_{q\rightarrow0^+}\frac{\psi_+'(m)}{iq}a_\text{1D}.
\end{split}
\end{equation}
We have use the following definition for the Fourier coefficients of any function $\Phi(R)$:
\begin{equation}
 \Phi(m)=\sqrt{\frac2a}\int_0^a\text{cos}(mkR)\Phi(R)dR \;.
\end{equation}
In the previous expressions $m,n=0,1,2,...,l_\text{max}$. Up to even odd effects, the solution $a_\text{CM}(l_\text{max})$ approaches exponentially fast its asymptotic value with increasing $l_\text{max}$, see Fig.~\ref{fig:scatt_exp}.

\paragraph{Two Particles in a Harmonic Potential.}

\begin{figure}[!b]
 \centering
 \includegraphics[width=0.82\linewidth]{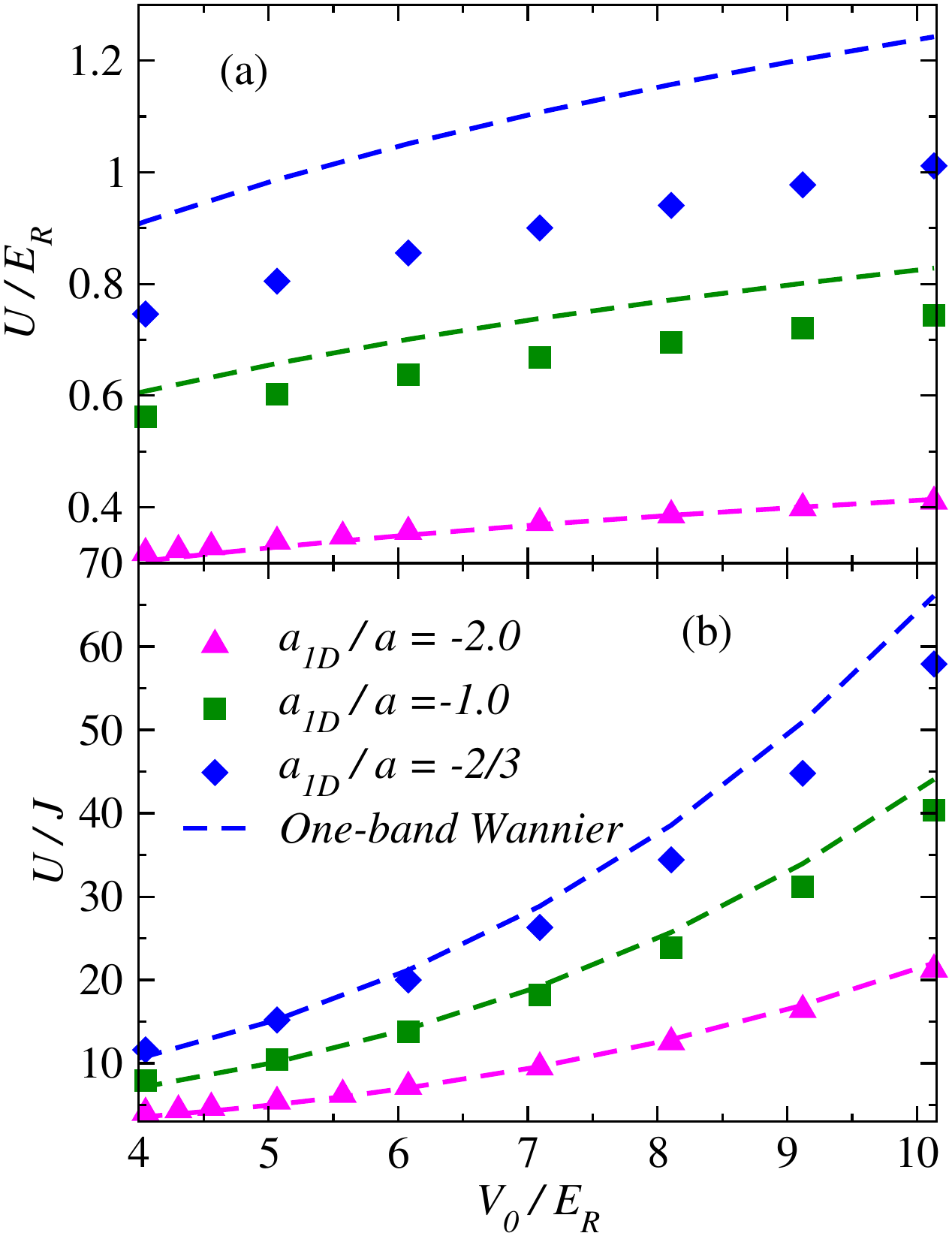}
 \vspace{-0.1cm}
 \caption{(Color online) (a) On-site repulsion $U$ computed using the ground state of two interacting particles in a harmonic potential (symbols), and one-band Wannier functions (lines). (b) $U/J$ obtained from the inverse confined scattering analysis (symbols) and using one-band Wannier functions (lines). }
 \label{fig:hubbard}
\end{figure}

We calculate the on-site repulsion $U$ by studying two $\delta$-interacting particles in a harmonic potential, which was exactly solved by Busch {\it et al.}~\cite{busch_98}. The two-body Hamiltonian can be decomposed in two independent Hamiltonians by introducing the center of mass and relative coordinates ($R,r$). Using $a_\text{HO}=\sqrt{\hbar/m\omega}$ and $\hbar\omega$ as the units of length and energy, respectively, one has
\begin{equation}
\begin{split}
&\tilde{H}_R=-\frac14\partial_{\tilde{R}}^2+\tilde{R}^2 \; , \\
&\tilde{H}_r=-\partial_{\tilde{r}}^2+\frac14\tilde{r}^2+\tilde{g}\delta(\tilde{r}) \;,
\end{split}
\end{equation}
where $\tilde{g}=-2a_\text{HO}/a_\text{1D}$. The eigenfunctions of $\tilde{H}_R$ are the harmonic oscillator eigenfunctions $\phi_n(\tilde{R})$. The Hamiltonian of the relative motion $\tilde{H}_r$ can also be solved analytically. The eigenfunctions with even parity are given by the Tricomi's confluent hypergeometric function $U(a,b,z)$
\begin{equation}
 \psi_n(\tilde{r})\varpropto e^{-\tilde{r}^2/4}U\left(\frac14-\frac{\tilde{E}_n}2,\frac12,\frac{\tilde{r}^2}2\right) \; .
\end{equation}
$\tilde{E}_n$ is the corresponding eigenenergy determined by the following relation
\begin{equation}
 -\frac{\tilde{g}}2=\frac{\Gamma\left(-\frac{\tilde{E}_n}2+\frac34\right)}{\Gamma\left(-\frac{\tilde{E}_n}2+\frac14\right)} \; ,
\end{equation}
where, $\Gamma(x)$ is the Gamma function.

The ground state wavefunction of the two interacting particles in the harmonic trap is then $\Psi_\text{GS}(\tilde{r},\tilde{R})=\phi_0(\tilde{R})\psi_0(\tilde{r})$, and the on-site repulsion reported in the main text was computed as: $U=\bra{\Psi_\text{GS}}H_\text{int}\ket{\Psi_\text{GS}}$. 

In Fig.~\ref{fig:hubbard}(a), we plot $U/E_R$ vs $V_0/E_R$ obtained as explained above and compare it to the results obtained using Wannier functions. In Fig.~\ref{fig:hubbard}(b), we plot the results for $U/J$ vs $V_0/E_R$ as obtained from the scattering analysis and also from Wannier functions. The plots in Fig.~\ref{fig:hubbard} make clear the importance of going beyond the traditionally used Wannier functions to compute $U$ and $U/J$ as the interaction strength increases.

\end{document}